# Designing Flexible GUI to Increase the Acceptance Rate of Product Data Management Systems in Industry

Zeeshan Ahmed [1, 2]
[1]Vienna University of Technology Austria,
[2]University of Wuerzburg Germany.

*Abstract*: Product Data Management (PDM) desktop and web based systems maintain the organizational technical and managerial data to increase the quality of products by improving the processes of development, business process flows, change management, product structure management, project tracking and resource planning. Though PDM is heavily benefiting industry but PDM community is facing a very serious unresolved issue in PDM system development with flexible and user friendly graphical user interface for efficient human machine communication. PDM systems offer different services and functionalities at a time but the graphical user interfaces of most of the PDM systems are not designed in a way that a user (especially a new user) can easily learn and use them. Targeting this issue, a thorough research was conducted in field of Human Computer Interaction; resultant data provides the information about graphical user interface development using rich internet applications. The accomplished goal of this research was to support the field of PDM with a proposition of a conceptual model for the implementation of a flexible web based graphical user interface. The proposed conceptual model was successfully designed into implementation model and a resultant prototype putting values to the field is now available. Describing the proposition in detail the main concept, implementation designs and developed prototype is also discussed in this paper. Moreover in the end, prototype is compared with respective functions of existing PDM systems .i.e., Windchill and CIM to evaluate its effectiveness against targeted challenge.

## 1. Introduction

Companies consist of different departments like management, marketing, accounts, production, quality and engineering etc. Every department has its own rules, regulations, data and staff. There is no doubt every department is important and expected to play a vital role in the progress of industrial enterprises but most important of all is the engineering or technical department, which is more responsible for the main product's development and production from all other departments. To successfully run the technical department hardware and software are deployed, processes are initiated and implemented, required number of technical staff is hired to produce the product under the implemented process using the provided resources. Problems initiate and start growing as a company grows due to the rapid increase in data with the lack of required in time information and project and resource management. As a result the company can face unnecessary additional increase in costs, delays in product completion, loss in quality and waste of time [7].

In the past, there were no such systems available to store, track and manage all the related product data. This doesn't mean that there was no system for data management; there were some systems to store the information about product, personnel involved in organization and financial details but there was no such comprehensive system to manage technical data. To cope with the problem of organizational technical data management a new system category was introduced i.e., Product Data Management (PDM). PDM is a digital way of maintaining engineering data of technical departments within organizations to improve the quality of products and processes. PDM products mainly manage information about design and manufacturing of products including technical operations and running projects.

Till now everything sounds perfect, but the problems initiate and start growing as company grows. These problems can happen because of lack in control over engineering processes, rapidly increasing data, lack of presence, lack of coordination among team members (staff), unclear product configurations, loss of experienced staff, conflicts between the central Information Systems (IS) organization, lack of suitable formal communications between departments, bureaucratic and complex engineering change control systems and lack of project and resource management. As the result company can face unnecessary additional increase in cost, delays in product completion, loss in quality and waste of time.

Successful PDM System Deployment in an organization (especially large one) is quite difficult because it is time consuming, expensive and most of the staff (belonging to corporate management, top level management, engineering management and other engineering and IT professionals) do not give importance to PDM System and without these person's support it is quite difficult to implement it. Moreover people don't want to involve in low level technical and business issues, don't want to spend money, look for fast pay back projects, don't have extra time, too much inertia in this company, lack of trust of users on management, job insecurity and incapable of handling PDM systems.

Some of the main reasons of lack in acknowledgement of PDM Systems in international market are some problematic issues and if these are resolved then it will be a great contribution to PDM System development, usage and marketing. The graphical user interfaces of most of PDM Systems are user unfriendly, nonflexible and slow (especially if the system is web based). In case, if PDM System is a client based application then the issue of platform independency is also there because in the new business models it is nearly impossible to mandate that all the potential users choose the same platform or the same operation system. Moreover PDM Systems normally deal with heavy amount of data but in most of the cases it is quite difficult to access or search needed information by using intelligent search mechanism. Without going into the details of all PDM System problems and residing within the scope of this research, focusing only one of the all current industrial unresolved issues in PDM System development i.e., unfriendly graphical user interface.

If a product is very productive and with lots of beneficial functionalities but if it is not easily usable then in most of the cases it becomes a flop in industry. Designing and implementing an intelligent and user friendly HMI for any kind of software or hardware application is always a challenging task for the designers and developers because it is very difficult to understand the psychology of the user, nature of the work and what best suits the environment. Normally PDM Systems offer many different services and functionalities at a time but the graphical user interfaces of most of the PDM Systems are not designed in a way that a user (especially a new user) can easily learn and use them. Most of the web GUI of PDM applications are with massive control implementation at user end, providing several options at a time which might not be in need of every user. Moreover the GUI of PDM applications are not flexible enough that a user can change the default orientation and placement of controls according to his need and choice.

The goal of this research is to support PDM with a web based platform independent PDM approach capable of providing flexible graphical user interface. This research is about to propose a new flexible web based GUI for multiple roles based client PDM systems capable of providing faster and better access of the system, options to change the default orientation of provided GUI controls, add or delete provided options, even a user can redesign a new graphical look by changing the default GUI according to his need and wish. Continue the discussion with the identification and detailed presentation of Problem Definitions in section 2. Then in section 3 and 4 Human Computer Interaction and Rich Internet Application are presented, as the part of state of the art. Newly proposed Approach towards some of the PDM System problems is presented in section 5 of this research paper. Narrowing the scope of this research paper and focusing only on the proposing of new flexible web based graphical user interface for PDM Systems, a new approach, its concept, implementation designs and developed Prototype is presented in sections 6, 7, 8 and 9 of this research paper. Later in section 10 of this research paper, the resultant prototype is compared with some existing PDM Systems, to evaluate its effectiveness. In the end, some existing limitations in developed prototype are presented in section 11 and discussion is concluded in section 12 of this research paper.

## 2. Problem Definition

PDM Systems offer different services and functionalities at a time for many types of roles/users like managers, designers, engineers etc. but the graphical user interface of most of the PDM Systems are not designed in a way that a user (especially a new user) can easily learn, use and adopt [6].

Even at times for the old users it becomes a massively complicated GUI with several options from which many of them are not even in use all the time. Moreover the GUI of PDM applications are not flexible enough that a user can change the default orientation of controls by redesigning the default GUI according to his need and choice, and can save it so that it can be reused later on.

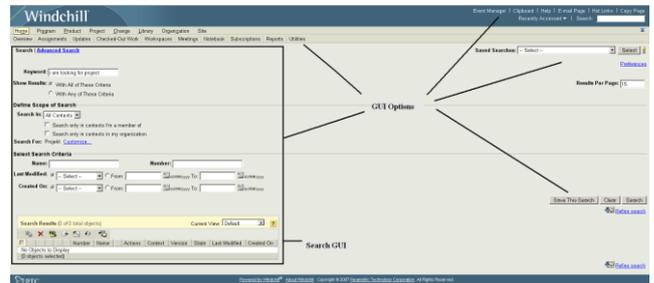

**Figure 1**. Windchill (Marked) Graphical User Interface

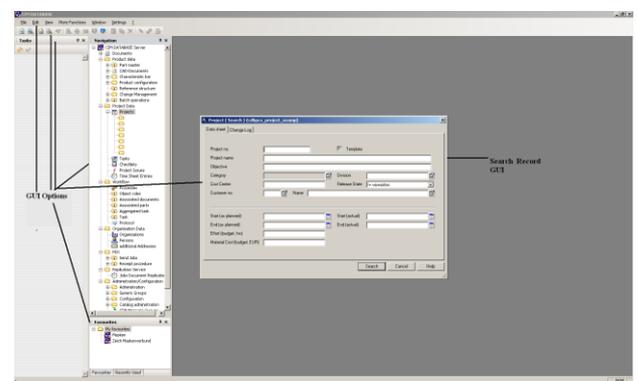

**Figure 2**. CIM Data Base (Marked) Graphical User Interface

PDM Systems are especially designed for the role based client users (multiple users playing different roles with different rights in the same organizations). So the probability of predicting that every user does not need all the options of the system all the time is very high. Moreover we can also say that the massive availability of all the controls to all the users all the time will also reduce the speed and efficiency in the work of the users because if a user will only be provided with some limited options with respect to his nature of job, rights and responsibilities then it will be much faster and more convenient for him to use and perform in the system. Moreover if the user is also provided with the flexibility of redesigning GUI by changing the default orientation of the provided controls (by adding or deleting) in default GUI of the PDM system according to his ease and the need then it will also be a useful contribution at the user's end. Currently available PDM Systems e.g. Windchill [1] as shown in Figure 1 and CIM Database [2] as shown in Figure 2 are excellent engineering data management systems but with not user friendly and flexible graphical user interface with structured input mechanism and massive input/output controls.

### 2.1 Example of Unfriendly GUI; Windchill
Windchill is a web enabled product data management application. It provide services in content and process management of organizational, technical and managerial data as shown in figure 1. Windchill is capable of decreasing product development time through efficient collaboration, reducing errors by automating processes and driving conformity to corporate standards, reducing scrap and rework by automatically sharing product data with downstream manufacturing systems and engineers, increasing efficiency by enabling engineers to quickly find and manage multiple forms of digital product content, eliminating mistakes associated with duplicate data, incomplete data, or manual data and technology risk by reducing the number of systems and databases to maintain and administer.

Windchill is heavily benefiting the industry by providing such an excellent engineering data management system at a time it is providing difficulties to its clients in using the system by providing a massive graphical user interface with more than 30 main GUI control options e.g. Program, Product, Project, Change, Library and Organization etc. and more than hundreds of sub main GUI control options at a time. Because of this massive provision of GUI controls it is difficult for the new user of Windchill to learn and adopt it quickly, and for the old users it is sometimes complex to search required main or sub main GUI options. Furthermore the GUI of the Windchill is not flexible so a user can not even make a little alteration according to his need and wish. Apart from the above discussion the currently available version of Windchill 9 is also very slow and this is another factor to be looked at and improved for better use.

### 2.2 Example of Unfriendly GUI; Windchill

CIM Database is a Client Server PDM with native client and an optional Web client to manage engineering data and support product creation and process implementation as shown in figure 2. CIM Database is a secure data management system for a range of centralized functions like selectable search, CAD systems, product and organizational data management and electronic data interchange (EDI). Likewise Windchill the CIM Database is also heavily benefiting the industry by managing engineering data and at the same time providing massive graphical user interface with more than 50 main GUI control options e.g. Product Data, Project Data, Workflow, Organization Data, PDX, Replication Services, Administration/Configurations and More Functions etc. and more than hundreds of sub main GUI control options at a time. But unlike Windchill the GUI of the desktop based client application of CIM Database is more user friendly and flexible.

As it is a desktop based client GUI, it provides some options in GUI control alignment and orientation e.g. user can change the placements of provided control trays, faster and with a better access to the controls etc. But apart from these advantages there is a big disadvantage of desktop based client GUI that it is not available using world wide web. The desktop based client must need to be installed before using this PDM System. Moreover using CIM Database a user is also restricted to perform only one task (while making search) at a time because CIM Database desktop based client search module is based on Single interface Data Input (SDI) concept. To overcome these deficiencies CIM has also launched a web based client GUI but with the almost same limitations of GUI earlier mentioned in Windchill's GUI discussion e.g. user unfriendliness and nonflexibility etc.

As discussed earlier and shown in figure 1 and 2 the GUI of Windchill and CIM Database is more or less same like traditional database applications consisting of several options like data manipulation forms to enter or edit or delete data, search forms to find needed information, print information, use of CAD for making designs etc. Moreover CIM Database and Windchill consists of massive (providing several options to each user which might not be needed every time but still they are there), nonflexible (GUI is not flexible enough; a user cannot change the orientation of controls according to his need and choice) and user unfriendly GUI (massive controls and non flexible GUI these are not much user friendly and it is quite difficult for a new user to adopt to them). Because of these deficiencies in the GUIs of the existing PDM Systems, a flexible web based GUI for the multiple roles based client PDM Systems is need to be proposed which should provide faster and better access of the System to the users by providing options to the users to change the default orientation of provided GUI controls according to the need and wish, better access to the controls, user's own choice look and feel which user can design, redesign, save, use and later can alter as well.

### 3. Human Computer Interaction

Targeting the challenge of proposition of designing a flexible web based graphical user interface development; I have chosen the field of Human Computer Interaction (HCI) to have complete understanding of graphical user interface design and development. HCI is the study of design, evaluation and implementation of interactive computing systems for human use [3]. Designing High quality HCI design is difficult to implement because of many reasons .i.e., market pressure of less time development, rapid functionality addition during development, excessive several iterations, competitive general purpose software and human behavior analysis.

Designing human computer interaction interface is an important and a complex task, but it could be simplified by decomposing task into subcomponents and maintaining relationships among those subcomponents. Task decomposition is a structured approach, applicable in both Software Engineering and Human Computer Interaction (HCI) fields depending on specific processes and design artifacts. Using design artifacts applications could be made for analysis and design by making the hand draw sketches to provide high level of logical design based on user requirements, usage scenarios and essential use cases. To design hand draw sketches there are some strategies to be followed .i.e., planning, sequential work flow, and levels of details.

### 3.1. HCI Design Principles

While evaluating or designing a user interface, it is important to keep in mind the HCI design principles. There are four major HCI design principles .i.e., Cooperation, Experimentation, Contextualization, Iteration and Empirical Measurement [4].

1. Cooperation plays a vital role in software project development. The most important and primitive principle of design process is the cooperation between both developers and the end users. Because in the design process with respect to the participatory design point of view there exists an uncommon principle .i.e., presenting the same issues with completely different perspectives and dimensions.
2. Generally experimentation is performed in the middle of recently acquired possibilities and the currently existing conditions. To assure that the present conditions are in conjunction with new ideas and supported by two primitive principles .i.e., concretization and contextualization of design, Principles are in associated with the above mentioned visions performing experiments with visions and hand on experience.
3. Design hooks its initial point with a particular configuration in which new computer based applications put into practice. Participatory design emphasizes on situations based on the implementation of iterative designs. The design composition of use is tied up with numerous social and technical issues. Generally participatory design of the development will specifically includes different kinds of participants i.e. Users, Managers and the design developers.
4. In design process, hang on to some issues which are not yet revealed, which are visioning the future product from design point of view and the construction of work from use point of view. But participatory design puts a controversial statement in accomplishing the same by making use of artifacts i.e. Prototype. Designers with cooperation will make use of the artifacts as a source for delegation of work. Participatory design also ends up with a controversial statement for trivial division of work in the process of development, which pleads overlap among the members of analysis, design and realization groups.
5. Empirical measurement is about to test the interface in early stages with the involvement of real users who come in contact with the interface on an everyday basis. Keep in mind that results may be altered if the performance level of the user is not an accurate depiction of the real human computer interaction. Furthermore its also about to establish quantitative usability specifics such as: the number of users performing the task, the time to complete the task and the number of errors made during the task.

### 3.2. Design Patterns

Like software engineering design patterns there are some graphical user interface design patterns i.e., Window Per Task, Direct Manipulation, Conversational Text, Selection, Form, Limited Selection Size, Ephemeral Feedback, Disabled Irrelevant Things, Supplementary Window and Step-by-Step Instructions. These patterns help designers in analyzing already designed graphical interfaces and designing a user friendly and required on demand graphical machine interface [3] e.g.

- Window per task helps in organizing the complete graphical user interface into different screens by providing the information about tasks per window screen.
- Direct Manipulation is a user machine interaction style where user interacts with system by directly using provided options.
- Conversational Text provides textual input information of designed interface's commands.
- Selection describes interaction style to choose options from provided list of options.
- Form describes discrete structures on screen.
- Limited Selection Size structures set of selections.
- Ephemeral Feedback provides the information about the natural flow of the interface.
- Disable Irrelevant Things guide in identifying and removing irrelevant interface elements.
- Supplementary Window provides information about supplementary windows.
- Step by Step Instructions help designer in sequencing set of actions

### 3.3. HCI Design Guidelines

A successful design interface can be implementable using the following guidelines .i.e.,

- Design mock ups should be implemented.
- Design should be presentable according to the need of the user.
- Criteria / principles should be applied to the design.
- Prepared according to the project proposal based on specified functional requirements.
- Should be evaluated with respect to the number of features asked to develop.
- Assessed by testing especially in work load conditions.
- Use case modeling should be used with the identification of user interface elements
- Should be flexible enough to adopt rapid prototype changes and modifications.
- Should be based on consistent sequences of actions required in similar situations.
- Should be based on identical terminologies used in prompts, menus, and help screens.
- Should be based on consistent color, layout, capitalization, fonts, and so on should be employed throughout.
- In case of massive GUI based many components, HCI should enable frequent users to use shortcuts o increase the pace of interaction with the use of abbreviations, special keys, hidden commands and macros.
- Provide informative feedback for every user action.
- Should categorized sequences of actions into groups.
- Should offer error prevention and simple error handling.
- Should provide permit easy and reversal of actions.
- Should reduce short term memory load
- The GUI should provide an obvious, intuitive, and consistent interface to the simulation system.

- The GUI should provide an efficient means for reusing component models.
- The GUI should provide different graphical layouts for different types of simulation applications.

## 4. Rich Internet Application

The term "Rich Internet Application" was introduced in a white paper of March 2002 by Macromedia. Rich Internet Applications (RIA) are web applications with features and functionalities of traditional desktop applications as well as web applications. Traditional web applications center all activities around client server architecture with a thin client where as RIA typically transfer the processing necessary for the user interface to the web client but keeps the bulk of the data (i.e., maintaining the state of the program) back on the application server.

RIA shares one characteristic with other web development technologies, an intermediate layer of code often called a Client Engine, between the user and the server. This client engine is usually downloaded as part of the instantiation of the application, and may be supplemented by further code downloads as use of the application progresses. The client engine acts as an extension of the browser, and usually takes over responsibility for rendering the application's user interface and for server communication. Using Client Engine RIA becomes richer, more responser, balanced, asynchronous and efficient.

- *Richness*: User interface behaviors are not obtainable using only HTML widgets available to standard browser based Web applications. This richer functionality may include anything that can be implemented in the technology being used on the client side, including drag and drop, using a slider to change data, calculations performed only by the client and not needing to be sent back to the server.
- *Responsively*: The interface behaviors are typically much more responsive than those of a standard Web browser that must always interact with a remote server. The most sophisticated examples of RIA is that it exhibits a look and feel of a desktop environment level. Using a client engine can also produce other performance benefits.
- *Balanced*: The demand for client and server computing resources is better balanced, so that the Web server needs not to be the working horse like in traditional Web application. This frees server resources and allows the same server hardware to handle more client sessions concurrently.
- *Asynchronous*: The client engine can interact with the server without waiting for the user to perform an interface action such as clicking on a button or link. This allows the user to view and interact with the page asynchronously from the client engine's communication with the server. This option allows RIA designers to move data between the client and the server without making the user wait. Perhaps the most common application of this is pre-fetching data, in which an application anticipates a future need for specific data and downloads it to the client before the user requests it, thereby speeding up a subsequent response. Google Maps use this technique to load adjacent map segments to the client before the user scrolls them into view.
- *Efficiency*: The network traffic may also be significantly reduced because an application-specific client engine can be more intelligent than a standard Web browser while deciding which data needs to be exchanged with servers. This can speed up the individual requests or responses because less data is being transferred for each interaction, and overall network load is reduced. However, over-use of asynchronous calls and pre-fetching techniques can neutralize or even reverse this potential benefit because the code cannot anticipate exactly what every user will do next, it is common for such techniques to download extra data, not all of which is actually needed, to many or all clients.

### 4.1. RIA Technologies

There are several RIA technologies available i.e., FLEX (Adobe), AJAX, OpenLaszlo and Silverlight (Microsoft).

- Flex is a highly productive, free open source framework for building and maintaining expressive web applications that deploy consistently on all major browsers, desktops, and operating systems. While Flex applications can be built using only the free Flex SDK, developers can use Adobe Flex Builder™ 3 software to dramatically accelerate development. Adobe Flex is a collection of technologies released by Adobe Systems for the development and deployment of cross platform rich Internet applications based on the proprietary Adobe Flash platform.
- AJAX is a free framework for quickly creating efficient and interactive Web applications that work across all popular browsers. AJAX stands for Asynchronous JavaScript and XML. AJAX is a type of programming which became popular in 2005 by Google. It is not a new programming language, but a new way to use existing standards. Its primary characteristic is the increased responsiveness and interactivity of web pages achieved by exchanging small amounts of data with the server "behind the scenes" so that entire web pages do not have to be reloaded each time, there is a need to fetch data from the server.
- OpenLaszlo is an open source platform for the development and delivery of rich Internet applications. It is released under the Open Source Initiative-certified Common Public License. Laszlo applications can be deployed as traditional Java servlets, which are compiled and returned to the browser dynamically. This method requires that the web server be running the OpenLaszlo server. OpenLaszlo was originally called the Laszlo Presentation Server (LPS).
- Microsoft Silverlight is a web browser plug-in that provides support for rich internet applications such as animation, vector graphics and audio-video playback. Silverlight provides a retained mode graphics system, similar to WPF and integrates multimedia, graphics, animations and interactivity into a single runtime. It is being designed to work in concert with XAML and is scriptable with JavaScript. XAML can be used for marking up the vector graphics and animations. Textual content created with Silverlight is more searchable and indexable than that created with Flash as it is not compiled, but represented as text (XAML). Silverlight can also be used to create Windows Sidebar gadgets.

**Table 1.** Comparison between RIA Technologies

| Content | Flex | Silverlight |
|---|---|---|
| **IDE GUI** | Yes | Yes |
| **Project User Interface declarations** | XML based (MXML) | XML based (MXML) |
| **Cross-platform** | Yes | Windows Only |
| **Server side integration** | object based, AMF | object based, AMF |
| **Worldwide usage** | Best | Poor |
| **Loading time / Boot** | Fast | Good |
| **3D** | HW supported | HW supported |
| **Components & Tools** | Better | Good |
| **Component integration with OS** | Good | Bad |

Based on the earlier discussed RIA based information and to conclude with one final technology for own flexible web based graphical user interface development, a comparison is performed between two most beneficial technologies of all i.e., Flex and Silverlight. As the result of comparison, on the basis of above presented results in table 1, Flex is chosen for the own flexible web based graphical user interface development for PDM Systems because Flex has the biggest advantage of being used for cross platform (operating system independent), having fast loading time and with provision of better tools and components.

## 5. Proposed Approach

Focusing on the need of an approach as the solution towards the problems of implementing a flexible graphical user interface for PDM System development, I have chosen and thoroughly investigated the field Human Computer Interaction, based on the resultant information of conduced research; a new approach has been proposed. As shown in Figure 3, the proposed approach consists of four different modules i.e. Flexible GUI, NLP Search, Data Manager and Data Representer.

Proposed approach is mainly for the development of a PDM system capable of providing a flexible web based graphical user interface, identifying user's structured and unstructured natural language based requests, processing natural language based user's requests to extract results from attached repositories [5], manage data in database management system and represent system outputted information as the result of user input in user's understandable format. In this research paper without going into the details of other three modules of proposed approach, will only discuss the module i.e. Flexible GUI.

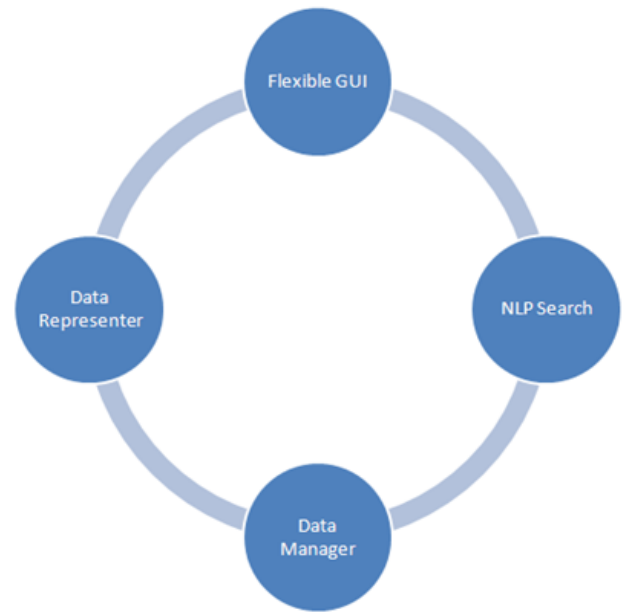

**Figure 3**. Conceptual Model of Proposed Approach

## 6. Proposed Flexible GUI

As shown in Figure 4, different kinds of users i.e., Businessman, Project Manager, Engineer and staff member etc. are need to interact with PDM System at a time. The major interests for a Businessman could be regarding the performance and quality of running projects. Project manager's job is to plan function including defining the project objective and developing a plan to accomplish the objective, organizing function involves identifying and securing necessary resources, determining tasks that must be completed, assigning the tasks, delegating authority, and motivating team members to work together on the project and manage running projects. Engineer is there to design product using CAD whereas other staff members could be involved in different tasks e.g. organization's personal and project data entrance and management etc. These different kinds of users have different kinds of psychologies to approach and use one PDM System. As PDM Systems consists of different options and these are designed and implemented for different kinds of users. The point to think is how a PDM system can provide a user system interaction mechanism which can satisfy all kinds of users because it is quite difficult to provide one graphical user interface which can satisfy all users by providing their needed components without creating a mess of options at GUI. Keeping this need in mind, proposed a new approach i.e. Flexible GUI, for PDM System development.

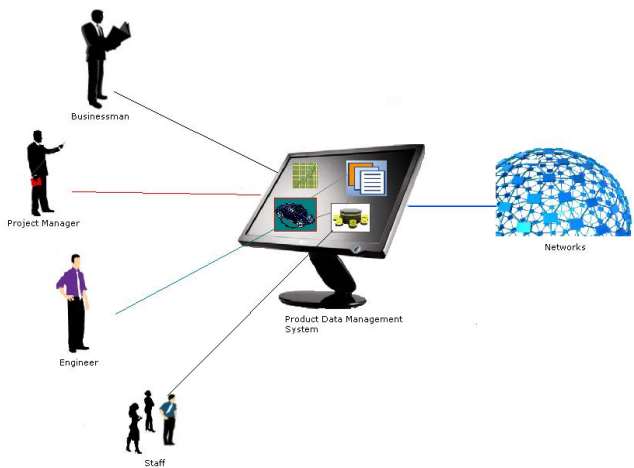

**Figure 4**. PDMs Multi Role based Users System Interactions

To manage the complex control flows necessary for GUIs designed for PDM Systems with flexible interfaces, a new way of Flexible GUI implementation is presented. Flexible GUI is a type of user interface that allows user to interact with the program in more ways than typing such as computers hand held devices and office equipment with images rather than text commands. Flexible GUI uses a combination of technologies and devices to provide a platform independent user interface which any user can interact with for the respective tasks. The design of Flexible GUI is based on three properties i.e. User friendliness, Model reusability and Application extensibility. User friendliness provides obvious, intuitive, and consistent interface, Model reusability provides an efficient means for reusing developed component and Application extensibility provides different graphical layouts for different types. Furthermore Flexible GUI's structure is flexible enough to accommodate graphical layout for different kind of user of different applications.

Flexible GUI is mainly a friendly web based graphical user interface proposed for product data management systems for better the human computer interaction. The main idea behind the proposition of a new web based graphical user interface is to improve user system communication by providing several options helping user by letting him change the default orientation of the GUI by changing the placements of provided controls, insertion of needed and deletion of unnecessary controls and redesigning completely new look and feel of the GUI, which is not at the moment possible in almost every PDM System.

Targeting the problem of a user friendly graphical user interface, the proposed flexible graphical user interface is designed keeping the need of provision of different services and functionalities at a time for many types of roles/user in mind. The proposed Flexible GUI for PDM applications is flexible enough that a user can change the default orientation of controls by redesigning the default GUI according to his need and choice, and can save it so that it can be reused later on because the GUI of most of PDM Systems is massively complicated with several different options at the same time to all the users from which many of them are not even in use of all the user at all the times. Furthermore the proposed Flexible GUI is especially designed for the multiple roles based clients providing faster and better access of the System.

## 7. Flexible GUI; Conceptual Designs

Following information obtained as the result of conducted research in the field of Human Computer Interaction, a mockup (draft physical sketch) of proposed Flexible GUI is designed for a prototype development of proposed approach. The mockup is presentable according to the need of the user, designed with respect to the criteria and principles followed by the system and flexible enough to adopt rapid prototype changes and modifications. The mockup is based on an interactive design displaying required quantitative material including images, windows and tools etc.

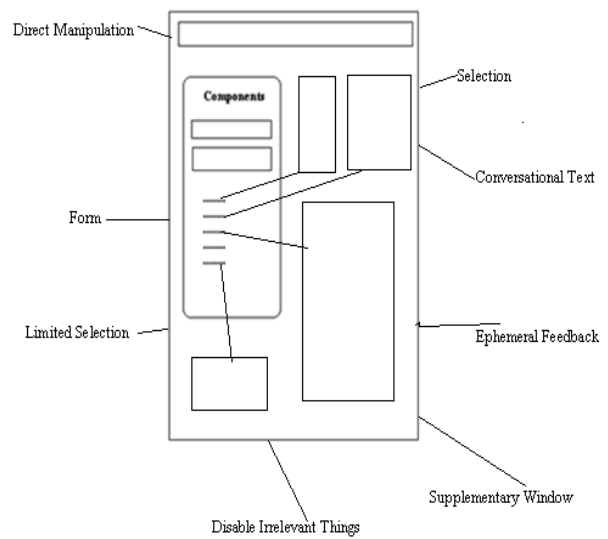

**Figure 5.** Mockup of Proposed Flexible GUI

The mockup of proposed Flexible GUI is needed to be implemented in the form of three different web pages using RIA technologies i.e. Components Page. The Component page is a prototype form of a flexible web based graphical user interface, consisting of a control container, giving an idea for placing all the components based options involved in the Product Data Management operations in a user's desired way by adding or deleting provided options. Further Component page also allows the user to redesign web based GUI with respect to its own choice by changing the GUI orientation by altering the GUI Component placements, changing the size of GUI control components (e.g. list boxes, mouse hover/click, drag drop, drop down list boxes, list boxes etc.) and changing the used color scheme, font, background etc. of in use GUI. The mockup of Component Page, as shown in Figure 5 is based on eight design patterns i.e. Direct Manipulation, Conversational Text, Selection, Form, Limited Selection, Ephemeral Feedback, Disable Irrelevant Things, Supplementary. Project implementation designs are created using these mockups for the prototype implementation of Flexible GUI using RIA technologies.

## 8. Flexible GUI; Implementation Designs

### 8.1. Design Methodology

Following three the classical tier application model, I have designed implementation methodology for the development of proposed prototype, as shown in Figure 6. The current version of proposed approach will be implanted with the use of Java (servlets and JSP) to handle user input, manage and retrieve data from the database. Tomcat is used as the main web server and middleware of the program. Users can access the web pages with the given URL and then can build graphical user interface or search the data after successful identity authorization. The data communication between three tiers is managed by Action Message Format (AMF) using the Simple Object Access Protocol (SOAP). AMF based client requests are delivered to the web server using Remote Procedure Call (RPC). The use of RPC allows presentation tier to directly access methods and classes at the server side. When data is request from user then a remote call is made from the user interface in the remote services' (via the server side includes) class members and the result is sent as an object of a Java class.

A web browser is mainly needed to access the developed application with a user of a specified universal resource link (URL). User will send a request to the web server through Hypertext Transfer Protocol (HTTP), the web server will pass the request to the application components. These application components are implemented using servlet/JSP, designed to handle user request coming from web server with the use of java remote classes. Then used servlets or JSP classe talks to the database server, perform the data transactions and send the response to the client. To increase flexibility of graphical user interface at client end, the development of front end is performed using Flex Flex (Builder 3 IDE), Relational database is designed and implemented using MySQL 5.

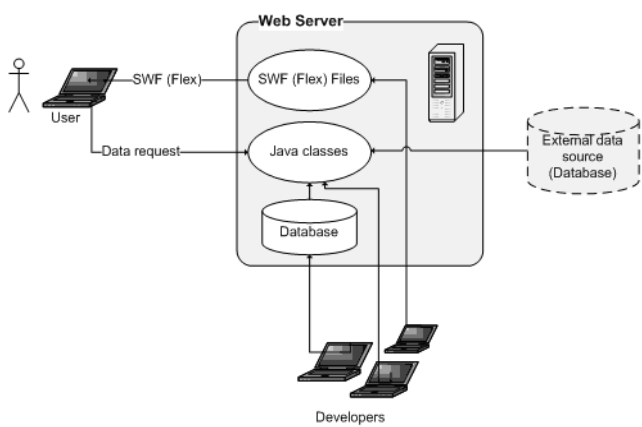

**Figure 6.** Implementation Design

### 8.2. System Sequence
As shown in figure 7, the Sequence design of the Flexible GUI consists of three components .i.e., Default GUI, Flexible GUI and Store GUI Setting.

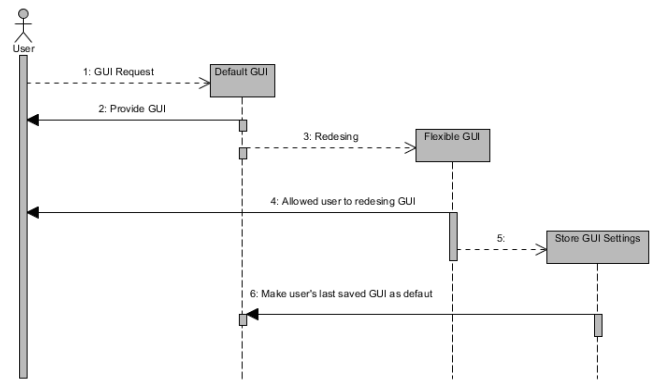

**Figure 7.** System Sequence Design

The job of Default GUI is to first identify user and then provide default system graphical user interface to the user, and incase a new graphical user interface is already designed and stored by user, then to provide his previously stored graphical user interface. Furthermore it also allows user to redesign a new graphical user interface with respect to this choice using providing components.

## 9. Flexibel GUI: Prototype

Following the constructed mockup, implementation designs, meeting the design requirements for a proposed Flexible GUI and residing with in the limited scope of this research's development, a prototype version of proposed approach is developed with the use of RIA technologies. This prototype version is Web application is capable of providing flexible graphical user interface with several different options (for multi role based clients) for Product Data Management Systems. The flexible web based graphical user interface is developed following designed mockup and divided into two sections as shown in Figures 8 and 9 i.e., Default Graphical Interface and User Graphical Interface.

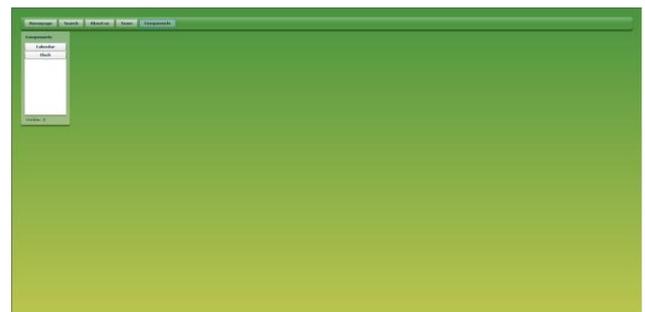

**Figure 8.** Prototype; Default Graphical Interface

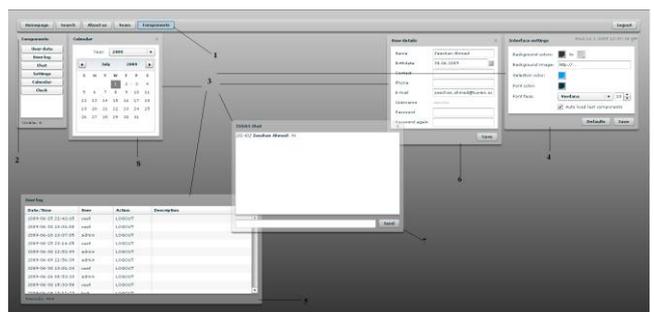

**Figure 9.** Prototype; User's Personal Graphical Interface

The default graphical interface is the graphical interface with some limited and basic options which can be accessed by the user and the guest. But the User graphical interface can only be accessed by the user after logging into the application with authenticated user name and password. User graphical interface is the actual interface presenting the prototype definition of proposed idea about Flexible GUI in conceptual design, as shown in figure 9 and described in table 2.

**Table 2.** Graphical Interface Page

| No | Option | Description |
|----|--------|-------------|
| 1 | Main Component Link | To enable the Graphical Interface |
| 2 | Components Tray | Providing all GUI options to the user to click and use |
| 3 | Components | All currently available components for PDM operations and GUI manipulations |
| 4 | Interface Setting | This components if used to change the outlook of the graphical interface by changing the color schemes and adding or removing image to the main interface |
| 5 | User Log | This components provides the detail of all the operations performed during the use of graphical interface, but this components if only visible to the user with administrative rights |
| 6 | User Details | This components provides the options to enter and alter user details |
| 7 | Chat | This component is providing option for in house chat to the login users to improve in house communications |
| 8 | Calendar | This is the simple calendar to enable user with date. |

This implemented prototype version is capable of
- Providing standard graphical interface designed by system.
- Providing flexible graphical user interface, so the user can redesign and reconfigure the interface itself to accommodate specific needs by Mouse Click and Drag Drop options.
- Providing several options to the user for GUI designing like user can change the look and feel by changing background colors, font and images, adding, deleting and altering components.
- Providing option to every user to save his own deigned GUI, so that the next time if the user comes online then he will be provided his own designed GUI rather than the default one. However he will still have the option to redesign or alter or restore the default GUI.

## 10. GUI Comparison

### 10.1. Prototype with CIM Database

The presented results in table 3 of performed comparison between the GUIs of prototype and CIM database demonstrates the contributions of prototype's GUI towards the PDM systems with respect to the scope, goal and earlier discussed defined problematic GUI based issues. The GUI of the client based desktop CIM database is quick and efficient in providing fast and easy access to provide the controls but at the same time it is not platform independent, it is not flexible enough so then a user can change the orientation of controls and can redesign GUI according to his choice and will but on the other hand the implemented prototype version of prototype is capable of providing these missing features in quick and efficient way.

**Table 3.** GUI Comparison between Prototype and CIM Database

| No | Jobs | CDB | Prototype |
|----|------|-----|-----------|
| 1 | Web based graphical user interface | No | Yes |
| 2 | Platform independent graphical user interface | No | Yes |
| 3 | Default GUI designed by system | Yes | Yes |
| 4 | User based Flexible Graphical Interface | No | Yes |
| 5 | The orientation of controls at GUI can be changed. | Yes | Yes |
| 6 | Reoriented controls of GUI can be saved and reused | No | Yes |
| 7 | Outlook of graphical user interface can be changed or newly designed. | No | Yes |
| 8 | Newly redesigned user based graphical interface can be saved and altered again. | No | Yes |
| 9 | Quick and efficient control's movement and data presentation. | Yes | Yes |

Prototype's GUI is platform independent, flexible enough so a user can redesign the default GUI by adding or deleting provided controls, changing the placements of in use controls, modifying the outlook of GUI according to his own choice and will and saving new redefined GUI for later reuse.

### 10.2. Prototype with Windchill

The presented results in table 4 of performed comparison between the GUIs of Prototype and Windchill database demonstrates the contributions of Prototype GUI towards the PDM systems with respect to the scope, goal and earlier discussed defined problematic GUI based issues. The GUI of the Windchill is web based platform independent application and with all needed options for engineering data management but at the same time if compared with Prototype's GUI then its GUI is slow, not flexible that a user can not change the orientation of controls and cannot redesign GUI according to his choice. Moreover in case of Windchill user is restricted to only use the default GUI with provided massive controls even when he is not in need of many of them. But in case of Prototype, the provided GUI is platform independent and flexible so that a user can redesign the default GUI by adding or deleting provided controls, changing the placements of in use controls, modifying the outlook of GUI according to his own choice and will and saving new redefined GUI for later reuse.

**Table 4.** GUI Comparison between Prototype and Windchill

| No | Jobs | CDB | Prototype |
|----|------|-----|-----------|
| 1 | Web based graphical user interface | Yes | Yes |
| 2 | Platform independent graphical user interface | Yes | Yes |
| 3 | Default GUI designed by system | Yes | Yes |

| 4 | User based Flexible Graphical Interface | No | Yes |
| 5 | The orientation of controls at GUI can be changed. | No | Yes |
| 6 | Reoriented controls of GUI can be saved and reused | No | Yes |
| 7 | Outlook of graphical user interface can be changed or newly designed. | No | Yes |
| 8 | Newly redesigned user based graphical interface can be saved and altered again. | No | Yes |
| 9 | Quick and efficient control's movement and data presentation. | No | Yes |

## 11. Limitations

The initial plan was to implement maximum possible PDM functionalities during the development of Flexible GUI of proposed approach but due the time limitations and limited scope of this research, development was restricted to the implementation of some of the functionalities putting some values but giving good idea that how can a complete Flexible GUI be implemented for a PDM System with all components and functionalities needed for a complete PDM System.

## 12. Conclusions

Targeting the challenge of proposition of web based flexible graphical user interface development; a thorough research has been conducted in Human Computer Interaction and RIA Technologies. Taking help from observed information from conducted research in respective field and using person research and development experience, I have proposed an approach. I have designed conceptual and implementation designs of proposed approach and implemented it using some software tools and technologies of present time i.e. Flex, Java, Antlr, MySQL, and presented developed prototype solutions.

In the end concluding the research and development efforts, we can say that proposed approach can put some values in enhancing PDM System development process by highlighting some existing challenges in PDM System development and proposing a new idea (along with conceptual and implementation designs) for flexible graphical user interface development to professional PDM System developing organization e.g. Windchill, CIM etc. The inclusive implementation of this proposed idea in PDM System development can put some values in increasing the market values of PDM Systems by increasing its acceptability in industry by improving its use amongst managerial, technical and office staff, because I strongly believe that if a product is very productive and with lots of beneficial functionalities (like PDM Systems) but not easily adoptable by its users then in most of the cases it becomes a failure in industry.

## 13. Acknowledgments

I am thankful to University of Wuerzburg Germany and Vienna University of Technology Austria for giving me the opportunity to keep working on this research project. I am thankful to Prof. Dr. Detlef Gerhard for his supervision during this research and pay my gratitude to Prof. Dr. Thomas Dandekar for his generous support. I also thanks to my beloved wife and colleague Mrs. Saman Majeed (Doctoral Researcher) for her support during this research, development and technical documentation.

## Supplementary Web Links

## Author Biographies

**Zeeshan Ahmed;** (born 15.01.1983) a Software Research Engineer by profession and presently working in the Department of Bioinformatics Biocenter University of Wuerzburg Germany. He has on record more than 12 years of University Education and more than 8 years of professional experience of working within different multinational organizations in the field of Computer Science with emphasis on software engineering of product line architecture based artificially intelligent systems.